\documentclass[epj]{svjour}
\usepackage{graphics}
\def\z0{\rm Z^0}
\newcommand{\as}{\alpha_{\rm s}}

\newcommand{\oaa}{{\cal O}(\as^2)}
\newcommand{\oaaa}{{\cal O}(\as^3)}
\newcommand{\oaaaa}{{\cal O}(\as^4)}
\newcommand{\epem}{\rm e^+\rm e^-}

\newcommand{\amz}{\as(M_{\rm Z^0})}

\def\mz{M_{\rm Z^0}}

\def\d2{D_2}
\def\oq{\char'134}
\def\lamsb{\Lambda_{\overline{MS}}}
\def\ecm{E_{cm}}

\def\m2{\mu^2}
\def\q{\rm q}

\def\q2{Q^2}
\def\asmu{\as (\mu^2)}
\def\asq{\as (\q2 )}

\def\R{\cal{R}}
\def\msbar{\overline{\mbox{MS}}}
\def\F{{\rm F}}
\def\mtau{\rm M_\tau}
\begin{document}
\hyphenation{extra-po-la-ted hadro-ni-sa-tion per-tur-ba-tive
non-per-tur-ba-tive}
\title{The 2009 World Average of $\as$ }
\author{Siegfried Bethke
}                     
\institute{MPI f\"ur Physik, F\"ohringer Ring 6, 
80805 Munich, Germany}
\date{Received: date / Revised version: date}
%
\abstract{
Measurements of $\as$, the coupling strength of the Strong Interaction
between quarks and gluons, are summarised
and an updated value of the world average of $\amz$ is derived. 
Building up on previous reviews, 
special emphasis is laid on the most recent
determinations of $\as$.
These are obtained
from $\tau$-decays, from global fits of electroweak precision 
data and from measurements of the proton structure function
$\F_2$, which are based 
on perturbative QCD calculations up to $\oaaaa$;
from hadronic event shapes and jet production in $\epem$
annihilation, based on $\oaaa$ QCD;
from jet production in deep inelastic scattering and from  
$\Upsilon$ decays, based on $\oaa$ QCD;
and from heavy quarkonia based on unquenched QCD lattice
calculations.
Applying pragmatic methods to deal with possibly underestimated errors
and/or unknown correlations,
the world average value of $\amz$
results in
$$\amz = 0.1184 \pm 0.0007\ .$$
The measured values of $\asq$, covering energy scales from
$Q \equiv \mtau = 1.78$\ GeV to 209~GeV, exactly follow the energy dependence
predicted by QCD and therefore significantly test the concept af 
Asymptotic Freedom.
}

\PACS{
      {12.38.Qk}{experimental tests of QCD}   
     } 
%
\maketitle
\section{Introduction}
\label{intro}
Quantum Chromodynamics (QCD) is the gauge field theory of the 
Strong Interaction \cite{qcd}.
QCD describes the interaction of quarks through the exchange of 
massless vector gauge bosons, the gluons, using concepts as known from
Quantum Electrodynamics, QED. 
QCD, however, is more complex than QED because
gluons themselves, other than photons in QED, carry the 
quantum-\oq~charge" of the Strong Interaction, such that gluons 
interact with each other.

As a consequence of the gluon self-coupling, 
QCD implies that the coupling strength $\as$, the analogue to the fine 
structure constant $\alpha$ in QED, becomes large 
at large distances or - equivalently - at low momentum transfers
\footnote{
\oq Large" distances $\Delta s$ correspond to
$\Delta s >$~1~fm, \oq low" momentum transfers to $Q < $~1~GeV/c.}.
Therefore QCD provides a qualitative reason for the observation
that quarks do not appear as free particles, but only exist as bound 
states of quarks, forming hadrons like protons, neutrons and pions.
Hadrons appear to be neutral w.r.t. the strong quantum charge.
The quark statistics of all known hadrons, their production cross sections
and decay times imply that there are three different states of the strong
charge.

In loose analogy to the behaviour of optical colours, 
the strong charge and the
corresponding new quantum number is called \oq colour charge".
Quarks carry one out of three different colour charges, while hadrons are
colourless bound states of 3 quarks or 3 antiquarks (\oq baryons"), or of a quark and an anti-quark (\oq mesons"). 
Gluons, in contrast to photons which do not carry
(electrical) charge by themselves, have two colour charges.

QCD does not predict the actual $value$ of $\as$, however 
it definitely predicts the functional form of the $energy$ $dependence$
of $\as$. 
While an increasingly large coupling at small energy scales leads to
the \oq confinement" of quarks and gluons inside hadrons,
the coupling becomes small at high-energy or short-distance reactions;
quarks and gluons are said to be \oq asymptotically free",
i.e. $\as \ \rightarrow$~0 for momentum transfers 
$Q\ \rightarrow \ \infty$.

The $value$ of $\as$, at a given
energy or momentum transfer scale\footnote{
Here and in the following, 
the speed of light and Planck's constant are set
to unity, $c = \hbar = 1$, such that energies, 
momenta and masses are given in units of GeV.} $Q$, 
must be obtained from experiment.
Determining $\as$ at a specific energy scale $Q$ is therefore a
fundamental measurement, to be compared with measurements of 
the electromagnetic coupling $\alpha$, of the
elementary electric charge, or of the gravitational constant.
Testing QCD, however, requires the measurement of $\as$ over 
ranges of energy scales: one measurement
fixes the free parameter, while the others test the specific QCD
prediction of confinement and of asymptotic freedom.

In this review the current status of
measurements of $\as$ is summarised. 
Theoretical basics of QCD and of the predicted energy 
dependence of $\as$ are given in Section~2.
Actual measurements of $\as$ are
presented in Section~3.
A global summary of these results and a determination 
of the world average value of $\amz$ are presented in Section~4.
Section~5 concludes and gives an outlook to
future requirements and developments.

\section{Theoretical basics}
\label{theory}

The concepts of Quantum Chromodynamics 
are presented in a variety of text books and articles, as e.g.
\cite{ellis-book,collins-book,yndurain-book,dissertori-book,
pdg,concise}, such that 
in the following, only
a brief summary of the basics of perturbative QCD and the running 
coupling parameter $\as$ will be given.

\subsection{Energy dependence of $\as$}
\label{running-as}

With the value of $\as$ known at a specific energy scale $Q^2$,
its energy dependence is given by the 
renormalisation group equation
\begin{equation} \label{eq-rge}
Q^2 \frac{\partial \asq}{\partial Q^2} = \beta \left( \asq \right) \ .
\end{equation}
\noindent The perturbative expansion of the $\beta$ function is calculated to
complete 4-loop approximation \cite{beta4loop}:
\begin{eqnarray} \label{eq-betafunction}
\beta (\asq ) = &-& \beta_0 \as^2(Q^2) - \beta_1 \as^3(Q^2) \nonumber \\
&-& \beta_2 \as^4(Q^2) -
\beta_3
\as^5(Q^2)  + {\cal O}(\as^6)\ ,
\end{eqnarray}
\noindent where
\begin{eqnarray} \label{eq-betas}
\beta_0 &=& \frac{33 - 2 N_f}{12 \pi}\ , \nonumber \\
\beta_1 &=& \frac{153 - 19 N_f}{24 \pi^2}\ , \nonumber \\
\beta_2 &=& \frac{77139 - 15099 N_f + 325 N_f^2}{3456 \pi^3}\ , \nonumber \\
\beta_3 &\approx & \frac{29243 - 6946.3 N_f + 405.089 N_f^2 + 1.49931 N_f^3}
        {256 \pi^4} 
\end{eqnarray}
\noindent and $N_f$ is the number of active quark flavours at the energy
scale $Q$.
The numerical constants in equation~\ref{eq-betas} are functions of the group constants $C_A = N$ and $C_F = (N^2 -1) / 2N$, for theories exhibiting $SU(N)$ symmetry.   
For QCD and $SU(3)$,
$C_A = 3$ and $C_F = 4/3$.

A solution of equation~\ref{eq-rge} in 1-loop approximation, i.e.
neglecting
$\beta_1$ and higher order terms, is
\begin{equation} \label{eq-as1loop}
\asq = \frac{\as (\mu^2 )}{1 + \as (\mu^2 ) \beta_0 \ln{\frac{\q2}{\mu^2}}}\ .
\end{equation}
\noindent 
Apart from giving a relation between the values of $\as$ at
two different energy scales, $\mu^2$ at which $\as$ is
assumed to be known, and $Q^2$ being another scale for which $\as$
is being predicted,
equation~\ref{eq-as1loop} also demonstrates the property of asymptotic freedom:
if $\q2$ becomes large and $\beta_0$ is positive, i.e. if
$N_f < 17$, $\asq$ will asymptotically decrease to zero
for $Q^2 \rightarrow \infty$.

Likewise, equation~\ref{eq-as1loop} indicates that $\asq$ grows to large values and diverges to infinity at small $Q^2$: for instance, with
$\as ( \mu^2 \equiv M^2_{\rm Z^0}) = 0.12$ and for 
typical values of $N_f = 2\ \dots \ 5$,
$\asq$ exceeds unity for $Q \leq \cal{O} \rm{(100~MeV \dots 1~GeV)}$.
Clearly, this is the region where perturbative expansions in $\as$ are not
meaningful anymore. 
Therefore energy scales below
the order of 1~GeV are regarded
as the nonperturbative region where confinement sets in, and
where equations~\ref{eq-rge} and~\ref{eq-as1loop} cannot be applied.

Including $\beta_1$ and higher order terms, similar but more complicated
relations for $\asq$, as a function of $\as (\mu^2 )$ and of
$\ln{\frac{\q2}{\mu^2}}$ as in equation~\ref{eq-as1loop}, emerge.
They can be solved numerically, such that for a given value of $\as (\mu^2 )$, 
choosing a suitable reference scale like the mass of the $\z0$ boson,
$\mu = M_{\z0}$,
$\asq$ can be accurately determined at any energy scale $\q2 \geq 1~\rm{GeV}^2$.

With
$$\Lambda^2 = \frac{\mu^2}{e^{1/\left( \beta_0 \as (\mu^2)\right) }}\ ,$$
\noindent a dimensional parameter $\Lambda$ is introduced  such that 
equation~\ref{eq-as1loop} transforms into
\begin{equation} \label{eq-as1loop-2}
\asq = \frac{1}{\beta_0 \ln (\q2 / \Lambda^2)}\ .
\end{equation}
\noindent 
Hence, the $\Lambda$ parameter is
technically identical to the energy scale $Q$ where $\asq$ diverges to
infinity.
To give a numerical example, $\Lambda \approx 0.1$~GeV for 
$\as (\mz \equiv 91.2\ \rm{GeV})$ = 0.12 and
$N_f$ = 5.

In complete 4-loop approximation and using the $\Lambda$-parametrisation, 
the running coupling is given \cite{alphas-4loop} by
\begin{eqnarray} \label{eq-as4loop}
\as (Q^2) &=& \frac{1}{\beta_0  L} 
               - \frac{1}{\beta_0^3 L^2} \beta_1 \ln  L  \nonumber \\
          &+& \frac{1}{\beta_0^3 L^3} \left( \frac{\beta_1^2}{\beta_0^2}
              \left( \ln^2 L - \ln L - 1 \right) + \frac{\beta_2}{\beta_0}
               \right) \nonumber \\
          &+& \frac{1}{\beta_0^4 L^4} \left( \frac{\beta_1^3}{\beta_0^3}
              \left( - \ln^3 L + \frac{5}{2} \ln^2 L + 2 \ln L - 
              \frac{1}{2}\right)\right)  \nonumber \\
          &-& \frac{1}{\beta_0^4 L^4} \left( 
               3 \frac{\beta_1 \beta_2}{\beta_0^2} \ln L
              + \frac{\beta_3}{2 \beta_0} \right)  
\end{eqnarray}
\noindent
where $L = \ln \left( Q^ 2 / \lamsb^2 \right) $.
The first line of equation~\ref{eq-as4loop} includes the 1- and the 2-loop
coefficients, the second line is the 3-loop and the third and the fourth 
lines denote the 4-loop correction, respectively.

\begin{figure}[ht]
\resizebox{0.49\textwidth}{!}{%
  \includegraphics{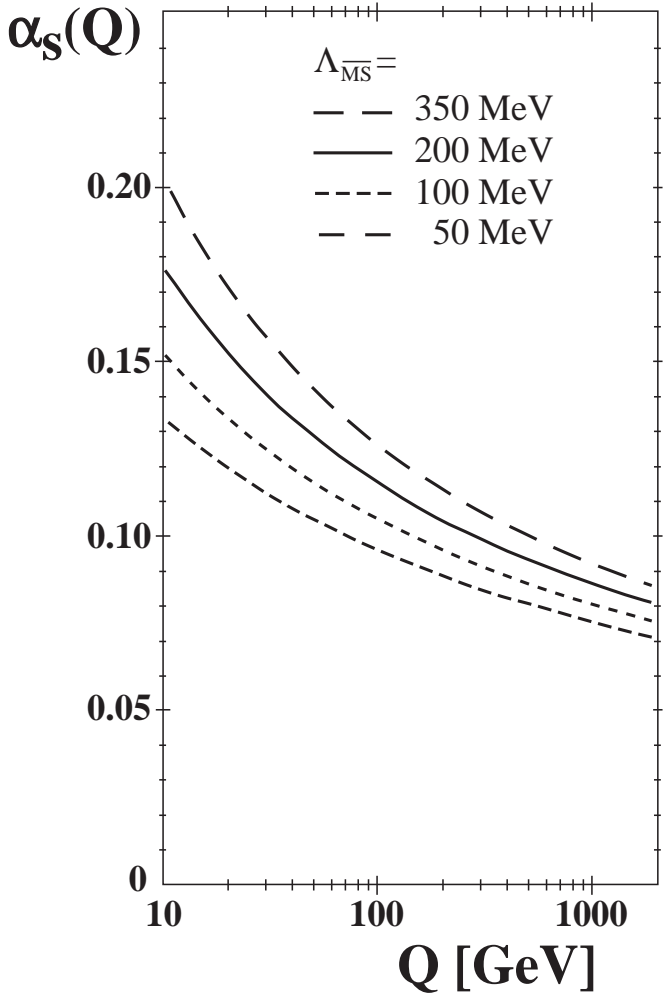}
}
\caption{The running of $\as (Q)$, according to
equation~\protect\ref{eq-as4loop},  in 4-loop approximation,
for different values of $\lamsb$.
\label{fig:asq-4L}}
\end{figure}

The functional form of $\as (Q)$, in 4-loop approximation and
for 4 different values of $\lamsb$, 
is diplayed in figure~\ref{fig:asq-4L}.
The slope and dependence on the actual value of $\lamsb$ 
is especially pronounced at small $\q2$, while at large $\q2$ both
the energy dependence and the dependence on $\lamsb$ becomes increasingly
feeble.

\begin{figure}[ht]
\resizebox{0.43\textwidth}{!}{%
  \includegraphics{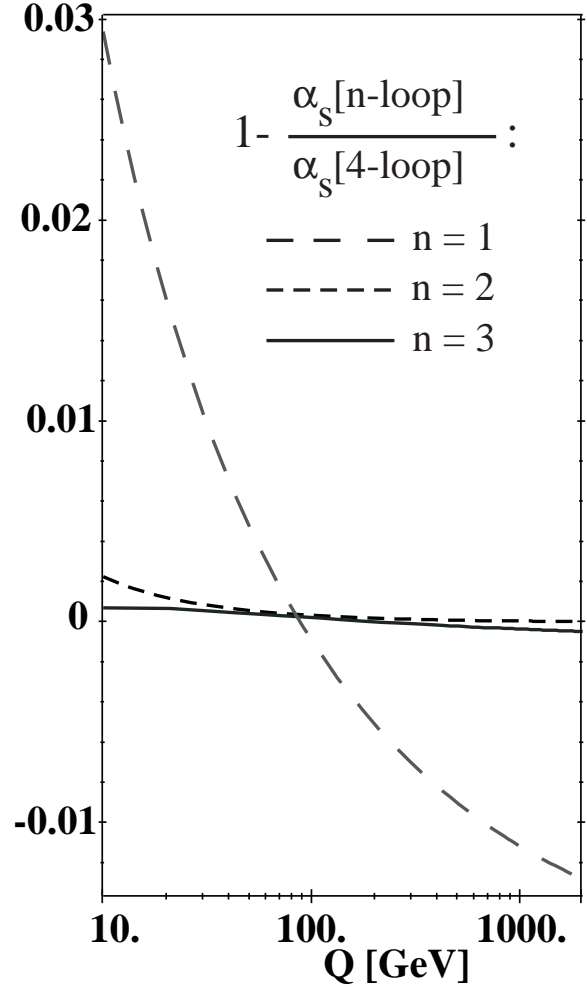}
}
\caption{Fractional difference between the 4-loop and the 1-, 2- and 3-loop
presentations of
$\as (Q)$, for $N_f = 5$ and $\lamsb$ chosen such that, in each order, $\amz
= 0.1184$.
\label{fig:asq-diff}}
\end{figure}

The relative size of higher order loop corrections and the degree of
convergence of the perturbative expansion of
$\as$ is 
demonstrated in figure~\ref{fig:asq-diff}, where
the fractional difference in the energy dependence of $\as$,
$(\as^{(4-loop)} - \as^{(n-loop)}) / \as^{(4-loop)}$, for $n$~=~1,~2 and~3, is presented.
The values of $\lamsb$ were
chosen such that $\amz = 0.1184$ in each order,
i.e., $\lamsb = 90$~MeV (1-loop), $\lamsb = 231$~MeV (2-loop), and
$\lamsb = 213$~MeV (3- and 4-loop).
Only the 1-loop approximation shows sizeable differences of up to
several per cent, in the energy and parameter range chosen, while the 2- and
3-loop approximations already reproduce the energy dependence of the 4-loop 
prediction quite accurately.

The parametrisation of the running coupling 
$\asq$ with $\Lambda$ instead of 
$\as (\mu^2)$ has become a common standard, see e.g.~\cite{pdg}. 
It will also be adopted here. 

\subsection{Quark threshold matching}
\label{matching}

Physical observables $\R$, when expressed as a function
of $\as$, must be continuous when crossing a quark threshold where
$N_f$ changes by one unit.
This determines that $\Lambda$ actually depends on the number of
active quark flavours.
$\Lambda$ will therefore be labelled $\lamsb^{(N_f)}$ to indicate these
peculiarities.
Also the slope of the energy dependence and, 
in approximations higher than 2-loop, the value 
of $\as$ change at the quark thresholds:

Construction of theoretical predictions which consistently match at a
quark flavour threshold leads to matching conditions for
the values of $\as$ above and below that threshold \cite{bernreuther}. 
In leading and in next-to-leading order, the matching condition is
$\as^{(N_f-1)} = \as^{N_f}$.
In higher orders, however, nontrivial matching
conditions apply \cite{alphas-4loop,bernreuther,larin}.
Formally these are, if the energy evolution of $\as$ is performed in 
$n^{th}$ order (or $n$ loops), of order ($n-1$).

The matching scale $\mu^{(N_f)}$ can be chosen in terms of the (running)
mass $ m_q ( \mu )$, or of
the constant, so-called pole mass $ M_q $.
For both cases, the relevant matching conditions are given in
\cite{alphas-4loop}. 
These expressions have a particluarly simple form for the 
choice\footnote{The results of reference~\cite{alphas-4loop} are also valid 
for other relations between $\mu^{(N_f)}$ and $m_q$ or $M_q$, as e.g. $\mu^{(N_f)}= 2 M_q$. 
For 3-loop matching, differences due to the freedom of 
this choice are negligible.}
$\mu^{(N_f)} = m_q (m_q)$ or $\mu^{(N_f)} = M_q$.
In this review, the latter choice will be used to perform 3-loop matching
at the heavy quark pole masses, in which case the matching condition reads, with
$a = \as^{(N_f)} / \pi$ and $a' = \as^{(N_f-1)} / \pi$:

\begin{equation} \label{Mq-matching}
\frac{a'}{a} = 1 + C_2\ a^2 + C_3\ a^3 \ ,
\end{equation}
\noindent
where $C_2 = - 0.291667$ and $C_3 = -5.32389 + (N_f-1)\cdot 0.26247$
\cite{alphas-4loop}.

\begin{figure}[ht]
\resizebox{0.49\textwidth}{!}{%
  \includegraphics{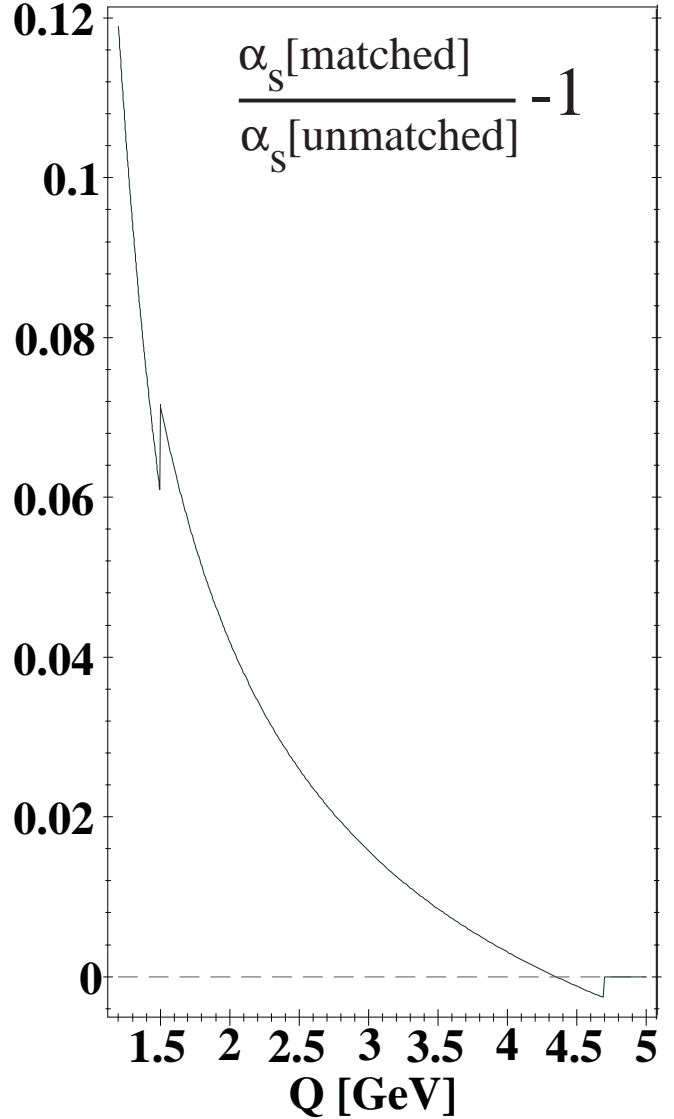}
}
\caption{The fractional difference between 4-loop running of 
$\as (Q)$ with 3-loop quark threshold matching
according to equations~\protect\ref{eq-as4loop} and~\protect\ref{Mq-matching}, with
$\lamsb^{(N_f = 5)}$~=~213~MeV and charm- and bottom-quark thresholds at the
pole masses,
$\mu^{(N_f = 4)}_c \equiv M_c = 1.5$~GeV and $\mu^{(N_f = 5)}_b \equiv M_b =
4.7$~GeV (full line),
and the unmatched 4-loop result (dashed line).
\label{fig:as-match}}
\end{figure}

The fractional difference of the
4-loop prediction for the running $\as$, using equation~\ref{eq-as4loop} with
$\lamsb^{(N_f = 5)}$~=~213~MeV and 3-loop matching at the charm- and 
bottom-quark pole masses,
$\mu^{(N_f = 4)}_c = M_c = 1.5$~GeV and $\mu^{(N_f = 5)}_b = M_b = 4.7$~GeV,
and the 4-loop prediction without applying matching and 
with $N_f$~=~5 throughout
are illustrated in figure~\ref{fig:as-match}.
Small discontinuities at the quark thresholds can be seen, such that
$\as^{(N_f-1)} < \as^{(N_f)}$ by about 2~per mille at the bottom- and about
1~per cent at the charm-quark threshold.
The corresponding values of $\lamsb$ are $\lamsb^{(N_f=4)} = 296$~MeV and $\lamsb^{(N_f=3)} = 338$~MeV.
In addition to the discontinuities, 
the matched calculation
shows a steeper rise towards smaller energies because of the larger 
values of
$\lamsb^{(N_f=4)}$ and $\lamsb^{(N_f=3)}$.
Note that the step function of $\as$ is not an effect
which can be measured; the steps are artifacts of the truncated
perturbation theory and the requirement that predictions for observables at
energy scales around the matching point must be consistent and independent of the
two possible choices of (neighbouring) values of $N_f$.

\subsection{Perturbative predictions of physical quantities}

In perturbative QCD, physical quantities ${\cal R}$
are usually given by a power series in $\as (\mu^2 )$, like
\begin{eqnarray} \label{eq-rseries}
{\cal R}(Q^2) &=&   P_{l} \sum_{n} R_n \as^n \nonumber \\
              &=& P_l \left( R_0 + R_1 \as (\mu^2) + R_2 (Q^2 / \mu^2 ) \as^2
(\mu^2 ) + ...\right)
\end{eqnarray}
\noindent 
where $R_n$ are the $n_{th}$ order coefficients of the perturbation series 
and $P_l R_0$ denotes the lowest-order value of $\cal R$.
$R_1$ is the
{\it leading order} (LO) coefficient,
$R_2$ is called the {\it next-to-leading order}
(NLO), $R_3$ is the {\it next-to-next-to-leading order}
(NNLO) and $R_4$ the N3LO coefficient.

QCD calculations in NLO perturbation theory are
available for many observables $\R$ in high energy particle reactions
like hadronic event shapes, jet production rates, scaling violations
of structure functions.
Calculations including the complete NNLO are
available for some totally inclusive quantities,
like the total hadronic cross section in $\epem \rightarrow\ hadrons$, 
moments and sum rules of structure functions in deep inelastic scattering 
processes,
the hadronic decay widths of the $\z0$ boson and of the $\tau$ lepton.
More recently, NNLO predictions were provided for exclusive
quantities like hadronic event shape distributions and differential
jet production rates in $\epem$ annihilation \cite{nnlo-shapes}, and
N3LO predictions for the hadronic width of
the $\z0$ boson and the $\tau$ lepton \cite{n3lo} became available.

A further approach to calculating higher order corrections is
based on the
resummation of logarithms which arise from soft and collinear
singularities in gluon emission \cite{resummation}.   
Application of resummation techniques and appropriate matching with fixed-order calculations are further detailed e.g. in \cite{concise}.

\subsection{Renormalisation}
\label{renormalisation}  

In quantum field theories like QCD and QED, 
physical quantities $\cal{R}$ 
can be expressed by a perturbation series in powers of the
coupling parameter $\as$ or $\alpha$, respectively.
If these couplings are sufficiently small, i.e. if
$\as \ll 1$, the series may converge sufficiently quickly such that it
provides a realistic prediction of $\cal{R}$ 
even if only a limited number of perturbative orders 
will be known. 

In QCD, examples of such quantities are cross sections, decay rates,
jet production rates or hadronic event shapes.
Consider $\cal{R}$ being dimensionless and
depending on $\as$ and on a single energy scale $Q$.
When calculating $\cal{R}$ as a perturbation series 
of a pointlike field theory in $\as$, ultraviolet
divergencies occur.
These divergencies are removed by the \oq renormalisation"
of a small set of physical parameters.
Fixing these parameters at a given scale and absorbing this way 
the ultraviolet divergencies, introduces a second but artificial
momentum or energy scale $\mu$.
As a consequence of this procedure, $\cal{R}$ and $\as$ become functions of
the renormalisation scale $\mu$.
Since $\cal{R}$ is  dimensionless, we assume that it only depends on the ratio
$Q^2 / \mu^2 $ and on the renormalized coupling $\as (\mu^2 )$:
$$ {\cal R} \equiv {\cal R}(Q^2 / \mu^2, \as );\ \as \equiv \as (\mu^2). $$

Because the choice of $\mu$ is arbitrary, however, 
the actual value of the experimental observable $\cal{R}$ cannot depend
on $\mu$, so that

\begin{equation} \label{eq-muindependence}
 \mu^2 \frac{{\rm d}}{{\rm d} \mu^2} {\cal R} (Q^2 / \mu^2 , \as )
= \left( \mu^2 \frac{\partial }{\partial \mu^2 } + \mu^2 \frac{\partial
\as}{\partial \mu^2} \frac{\partial }{\partial \as } \right) {\cal R} 
=^{\hskip -5pt !}\ 0 \ ,
\end{equation}

\noindent where the derivative is multiplied with $\mu^2$
in order to keep the expression dimensionless. 
Equation~\ref{eq-muindependence} implies that any explicite dependence of
$\cal{R}$ on
$\mu$ must be cancelled by an appropriate $\mu$-dependence of $\as$
to all orders.
It would therefore be natural to identify the renormalisation scale with 
the physical energy scale of the process, $\mu^2 = Q^2$, eliminating the
uncomfortable presence of a second and unspecified scale. 
In this case, $\as$ transforms to the \oq running coupling constant"
$\asq$, and the energy dependence of $\cal{R}$ enters only 
through the energy dependence of $\asq$.

The principal independence of a physical observable $\R$
from the choice of the renormalisation scale $\mu$ was expressed in
equation~\ref{eq-muindependence}.
Replacing $\as$ by $\as (\mu^2)$, using
equation~\ref{eq-rge}, and inserting the perturbative expansion of $\R$ 
(equation~\ref{eq-rseries}) into equation~\ref{eq-muindependence} results, 
for processes with constant $P_l$, in
\begin{eqnarray} \label{eq-muindependence2}
0 = \mu^2 \frac{\partial R_0}{\partial \mu^2} 
    &+& \as (\mu^2) \mu^2 \frac{\partial R_1}{\partial \mu^2} \nonumber 
    + \as^2 (\mu^2) \left[ \mu^2 \frac{\partial R_2}{\partial \mu^2} -
    R_1 \beta_0 \right]  \nonumber \\
    &+& \as^3 (\mu^2) \left[ \mu^2 \frac{\partial R_3}{\partial \mu^2} -
    [R_1 \beta_1 + 2 R_2 \beta_0] \right] \nonumber \\
    &+& {\cal O} (\as^4) \ .
\end{eqnarray}
Solving this relation requires that the coefficients of $\as^n (\mu^2)$ vanish
for each order $n$.
With an appropriate choice of integration limits one thus obtains
\begin{eqnarray} \label{eq-R-mudependence}
R_0 &=& {\rm const.}\ , \nonumber \\
R_1 &=& {\rm const.}\ , \nonumber \\
R_2 \left(\frac{Q^2}{\mu^2}\right) &=& R_2 (1) - \beta_0 R_1 \ln
\frac{Q^2}{\mu^2}\ , \nonumber \\
R_3 \left(\frac{Q^2}{\mu^2}\right) &=& R_3 (1) - 
[ 2 R_2(1) \beta_0 + R_1 \beta_1
]  \ln \frac{Q^2}{\mu^2} \nonumber \\
&+& R_1\beta_0^2 \ln^2 \frac{Q^2}{\mu^2}
\end{eqnarray}
\noindent
as a solution of equation~\ref{eq-muindependence2}.

Invariance of the complete perturbation series
against the choice of the renormalisation scale $\mu^2$ 
therefore implies that the
coefficients
$R_n$, except $R_0$ and $R_1$, explicitly depend on $\mu^2$.
In infinite order, the renormalisation scale dependence of $\as$ and of the
coefficients $R_n$ cancel; in any finite (truncated) order, however, the
cancellation is not perfect, such that all realistic perturbative QCD
predictions include an explicit dependence on the choice of the
renormalisation scale.

The scale dependence is most pronounced in leading order QCD
because $R_1$ does not explicitly
depend on $\mu$ and thus, there is no cancellation of
the (logarithmic) scale dependence of $\as (\mu^2)$ at all.
Only in next-to-leading and higher orders, the scale dependence of the
coefficients $R_n$, for $n \ge 2$, partly cancels that of $\as (\mu^2)$.
In general, the degree of cancellation improves with the inclusion of 
higher orders in the perturbation series of $\R$.

Renormalisation scale dependence 
is often used to test and specify uncertainties
of theoretical calculations  
for physical observables.
In most studies, the central value
of $\asmu$ is determined or taken for $\mu$ equalling the 
typical energy of the underlying scattering reaction, like e.g.
$\mu^2 = \ecm^2$ in $\epem$ annihilation. 
Changes of the result when varying this
definition of $\mu$ within \oq reasonable ranges"
are taken as systematic higher order
uncertainties. 

There are several proposals of how to optimise or fix the
renormalisation scale, see e.g. \cite{stevenson,grunberg,blm,siggiscale}
Unfortunately, there is no common agreement of how to optimise
the choice of scales or how to
define the size of the corresponding uncertainties.
This unfortunate situation should be kept in mind when 
comparing and summarising results from
different analyses.

In next-to-leading order, variation of the renormalisation scale is 
sufficient to assess and include theoretical uncertainties due to the
chosen
renormalisation scheme and the limited (truncated) perturbation
series.
In NNLO and higher, howerer, $both$ the renormalisation scale and the
renormalisation scheme should be varied for a complete assessment.
While it has become customary to include renormalisation scale
variations when applying theoretical predictions, changes of the
renormalisation scheme are rarely explored. 
Instead, 
the so-called \oq modified minimal
subtraction scheme" ($\msbar$)
\cite{msbar} is commonly used in most analyses, which is also the 
standard choice in this review.

\subsection{Nonperturbative methods}

At large distances or low momentum transfers, $\as$ becomes large and
application of perturbation theory becomes inappropriate.
Nonperturbative methods have therefore been developed to describe
strong interaction processes at low energy scales of typically
$Q^2 < 1$~GeV$^2$, such as the fragmentation of quarks and gluons
into hadrons (\oq hadronisation") and 
the masses and mass splittings of mesons. 

{\em Hadronisation models} are used in Monte Carlo approaches
to describe the transition of quarks and gluons into hadrons.
They are based on QCD-inspired mechanisms like the
\oq string fragmentation" \cite{string,pythia} or 
\oq cluster fragmentation" \cite{herwig}, and are usually 
implemeted, together with perturbative QCD shower and/or
(N)LO QCD generators, in models describing complete
hadronic final states in high energy particle collisions.
Those models contain a number of free parameters 
which must be adjusted in order to reproduce experimental data
well.
They are indispensable tools not only for detailed QCD studies
of high energy collision reactions, but are also important to 
assess the resolution and acceptance of large particle detector systems.

{\em Power corrections} are an analytic approach
to approximate nonperturbative hadronisation effects by means of
perturbative methods, introducing a universal, non- perturbative parameter
$$\alpha_0 (\mu_I ) = \frac{1}{\mu_I } \int^{\mu_I}_0 {\rm d} k\ \as (k) $$
to parametrise the unknown behaviour of $\as (Q)$ below a certain
infrared matching scale $\mu_I $ \cite{powcor}.
Power corrections are regarded as an alternative approach to describe
hadronisation effects on event shape distributions, instead of using
phenomenological hadronisation models.

{\em Lattice Gauge Theory} is one of the most developed nonperturbative 
methods (see e.g. \cite{weisz}) and is used to calculate, for instance, 
hadron masses, mass splittings and QCD matrix elements.
In Lattice QCD, field operators are applied on a discrete, 4-dimensional
Euclidean space-time of hypercubes with side length $a$.
Finite size lattice and spacing effects are studied by using increasing lattice sizes and decreasing lattice spacing $a$, hoping to eventually 
approach the continuum limit.
With ever increasing computing power and refined Monte Carlo methods,
these calculations significantly matured over time and recently provided 
predictions of the proton (and other hadron) masses to better than 2\%
\cite{fodor}, and determinations of $\as$ from quarkonia mass splittings
with a precision of better than 1 \% \cite{davies08}.

\section{Measurements of $\as$}

Since almost 30 years, determinations of $\as$ continue to be
at the forefront of experimental studies and tests of QCD.
Increasing precision of QCD predictions and methods, improved 
understanding and parametrisation of non- perturbative effects,
increased data quality and statistics and the availabilty of 
data over large ranges of energy and from a large variety of
processes have led to an ever increasing precision and depth
of these studies. 
The timely development of $\as$ determinations was documented and 
summarised in a number of summary articles, see e.g.
\cite{altarelli,concise,pdg,sb-06}, on which this review is
intended to build up. 
Since about the year 2000, the precision of $\as$ determinations
and the multitude of results from various processes and ranges
of energies provided experimental proof 
\cite{concise,sb-06} of the 
concept of asymptotic freedom.

In this review, special emphasis is laid on an update of the
review from 2006 \cite{sb-06}, concentrating on the most
recent results which are mostly based on further improved 
theoretical predictions and/or experimental precision:

\begin{itemize}
\item 
perturbative QCD predictions in complete N3LO 
for the hadronic widths of the
$\z0$ boson and the $\tau$ lepton are now available, 
improving further the completeness of the perturbative series and
providing increased control of remaining theoretical uncertainties;

\item
improved lattice QCD simulations with vacuum polarisation from u, d and s
quarks, updating previous determinations of $\as$ and quoting 
overall  uncertainties of less than 1\%;

\item
an improved extraction of $\as$ from radiative decays of
the $\Upsilon$(1s);

\item
a combined analysis of non-singlet structure functions from deep inelastic scattering data,  based on QCD predictions complete to N3LO;

\item
a combined analysis of inclusive jet cross section measurements
in neutral current deep inelastic scattering at high $Q^2$;
 
\item
determinations of $\as$ from hadronic event shapes and jet rates 
in $\epem$ annihilation final states, an important and (experimentally) very precise environment, based on the new and long awaited
QCD predictions in complete NNLO QCD.
 
\end{itemize}

These recent results are superior to and thus replace
a large number of $\as$ 
determinations published before 2006 and summarised in \cite{sb-06}.
While those previous measurements still remain valid, only the new 
results listed above are presented and summarised below.

\subsection{$\as$ from $\tau$-lepton decays}

Determination of $\as$ from $\tau$ lepton decays 
is one of the most actively studied fields to measure this basic
quantity.
The small effective energy scale, $Q = M_\tau =
1.78$~GeV, small nonperturbative contributions to 
experimental measurements of a total inclusive observable,
the normalised hadronic branching fraction of $\tau$ lepton decays,
\begin{equation}
R_{\tau} = \frac{\Gamma (\tau^- \rightarrow {\rm hadrons}\ \nu_{\tau})} {\Gamma (\tau^-
\rightarrow {\rm e^-} \overline{\nu}_e \nu_{\tau})}\ , \nonumber
\end{equation}
invariant mass distributions (spectral functions) of
hadronic final states of $\tau$-decays,
and the \oq shrinking error" effect of the QCD energy evolution 
of $\as$ towards higher energies 
\footnote{
According to equations~1 and~2, in leading order,
$\Delta \as (Q^2)/\as (Q^2) \sim \as (Q^2)$.
Therefore, since $\as (Q^2)$ decreases by about
a factor of 3 when running from $Q^2 = M^2_\tau$ to $M^2_Z$,
the relative error of $\as$ also 
decreases by about a factor of 3.}
provide the means for
one of the most precise determinations of $\amz$.
Theoretically,
$R_\tau$ is predicted to be~\cite{braaten}
\begin{equation} 
R_\tau = N_c \ S_{EW} \left| V_{ud} \right|^2
( 1 + \delta'_{EW} + \delta_{\rm pert} + \delta_{\rm nonpert})\ .
\end{equation}
Here, $S_{EW} = 1.0189 (6)$ \cite{ew1} and $\delta'_{EW} = 0.001 (1)$
\cite{ew2} are electroweak corrections, 
$\left| V_{ud} \right|^2 = 0.97418 (27)$ \cite{pdg}, 
$\delta_{\rm pert}$ and $\delta_{\rm nonpert}$ are perturbative and
nonperturbative QCD corrections. 
Most recently, $\delta_{\rm pert}$ was calculated to
complete N3LO perturbative order, $\oaaaa$ \cite{n3lo};
it is of similar structure as the one for the hadronic branching
fraction $R_Z$ of the $Z^0$ boson. 
Based on the operator product expansion (OPE) \cite{ope},
the nonperturbative corrections are estimated to be small~\cite{braaten},
$\delta_{\rm nonpert} \sim -0.007 \pm 0.004$.
A comprehensive review of the physics of hadronic $\tau$ decays 
was given in \cite{davier06}.

\begin{figure}
\resizebox{0.49\textwidth}{!}{%
  \includegraphics{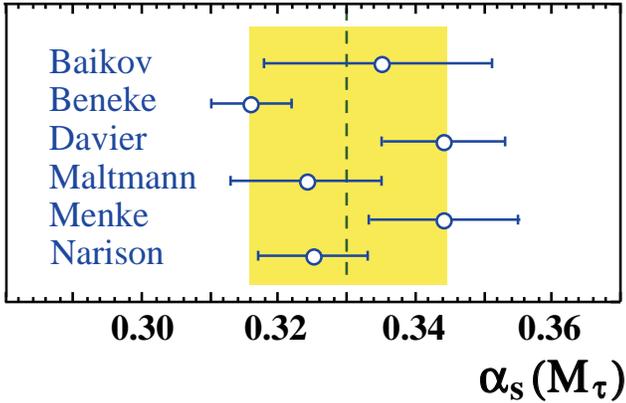}
}
\caption{Determinations of $\as$ from hadronic $\tau$ lepton
decays \cite{n3lo,beneke,davier08,maltman,menke,narison}. 
The results are all based on the
same experimental data and on perturbative QCD predictions to $\oaaaa$, 
however vary in preference or range of the perturbative expansion 
and inclusion and treatment of nonperturbative corrections.
The vertical line and shaded band show the average value and
uncertainty used as overall result from $\tau$ decays in this
review.}
\label{fig:asmtau-09}       
\end{figure}

Recently, several authors have revisited the determination
of $\as$ from $\tau$ decays
\cite{n3lo,beneke,davier08,maltman,menke,narison}.
These studies are based on data from LEP \cite{lep-tau}
and - partly- from BABAR \cite{babar}. 
They differ, however, in the detailed treatment and usage of 
the perturbative QCD expansion of $R_{\tau}$.
In particular, the usage of either fixed order (FOPT) 
or contour improved perturbative expansion (CIPT),
and differences in the treatment and inclusion  
of nonperturbative corrections, leads to 
systematic differences in the central values of $\as (M_\tau)$,
ranging from 0.316 to 0.344, as summarised in 
figure \ref{fig:asmtau-09}.
Results based on FOPT turn out to be systematically lower than 
those using CIPT -
a trend being known for quite some time, and being actively disputed in
the literature, but not finally being solved.

The results shown in figure~\ref{fig:asmtau-09}, within their assigned total
uncertainties, are partly incompatible with each other.
This is especially true if considering that they are based
on the same data sets.
The main reason for these discrepancies is the usage of either the
FOPT or the CIPT pertubative expansions.
Note that only the result of Baikov et al. 
\cite{n3lo} averages between
these two expansions, and assigns an
overall error which includes the difference between these two.

In view of these differences and for the sake of including
the apparent span between different perturbative
expansions in the overall error, the range shown as shaded band
in figure \ref{fig:asmtau-09} and the corresponding
central value is taken as the
final result from $\tau$-decays, leading to 
$$\as (M_\tau) = 0.330 \pm 0.014\ .$$
Running this value to the $Z^0$ rest mass of 91.2 Ge,V using the 4-loop
solution of the $\beta$-function (equation~6) with 3-loop
matching at the heavy quark pole masses $M_c = 1.5$ GeV and $M_b = 4.7$
GeV, results in
$$ \amz = 0.1197 \pm 0.0016\ .$$
This value will be included in determining the world average
of $\amz$ as described in section~4.

\subsection{$\as$ from heavy quarkonia}

Heavy Quarkonia, i.e. meson states consisting of
a heavy (charm- or bottom-) quark and antiquark, 
are a classical testing ground for QCD, see e.g. \cite{hq-review}.
Masses, mass splittings between various states, and decay rates
are observables which can be measured quite accurately, and which can be
predicted by QCD based on both perturbation theory and on 
lattice calculations.

\subsubsection{$\as$ from radiative $\Upsilon$ decays}

Bound states of a bottom quark and antiquark 
are potentially very sensitive to the value of $\as$ because the hadronic 
decay proceeds via three gluons, $\Upsilon \rightarrow ggg 
\rightarrow$~hadrons.
The lowest order QCD term (i.e. $P_l$ in
equation~\ref{eq-rseries}) 
for the hadronic $\Upsilon$ decay width
already contains $\as$ to the $3^{rd}$ power.
The situation is more complicated, however, due to 
relativistic corrections and to the
unknown wave function of the $\Upsilon$ at the origin.

The wave function and relativistic corrections largely cancel out in
ratios of decay widths like
$$ R_{\gamma} = \frac{\Gamma (\Upsilon \rightarrow \gamma gg)}
   {\Gamma (\Upsilon \rightarrow ggg)} $$
which therefore are the classical observables for precise
determinations of $\as$.

In \cite{brambilla}, recent CLEO data \cite{cleo-U} are used to 
determine $\as$ from radiative decays of the $\Upsilon$(1S).
The theoretical predictions include QCD up to NLO.
They are based 
on recent estimates of colour octet operators and avoid any model
dependences.
The value obtained from this study is
$$\amz = 0.119^{+0.006}_{-0.005}\ . $$
It is compatible with previous results from similar studies,
see e.g. \cite{concise,sb-06} and references quoted therein.
It will be included in the calculation of the new world average of
$\amz$.

\subsubsection{$\as$ from lattice QCD}

Determinations of $\as$ based on lattice QCD calculations 
have become increasingly inclusive and precise in the past,
including light quarks (u,d and s) in the vacuum polarisation
and incorporating finer lattice spacing.

In a recent study by the HPQCD collaboration \cite{davies08},
the QCD parameters - the bare coupling constant and the
bare quark masses - are tuned to reproduce the measured 
$\Upsilon$' - $\Upsilon$ meson mass difference.
The u, d and s quark masses are adjusted to give correct values of
various light meson masses.
With these parameters set, there are no other free physical parameters, 
and the simulation is used to provide accurate QCD predictions.
Nonperturbative values  of several short-distance quantities are 
computed and compared to respective perturbative calculations
which are given in NNLO perturbation theory.
From a fit to 22 short distance quantities, the value of
$$\amz = 0.1183 \pm 0.0008$$
is finally obtained.
The total error incudes 
finite lattice spacing, finite lattice volume, perturbative and extrapolation uncertainties. 
This result will be an important ingredient of the new world average
determined in section~4.

\subsection{$\as$ from deep inlastic scattering}

Measurements of scaling violations
in deep inelastic lepton-nucleon scattering belong
to the earliest methods used to determine $\as$.
The first significant determinations of $\as$, being
based on perturbative QCD prediction in NLO,
date back to 1979 \cite{yndurain-79}.

Today, a large number of results is available 
from data in the energy ($Q^2$)
range of a few to several thousand GeV$^2$, using electron-, muon- and
neutrino-beams on various fixed target materials, as
well as electron-proton or positron-proton colliding beams at HERA. 
In addition to scaling violations of structure functions, $\as$ is also
determined from moments of structure functions, from QCD sum rules and --- 
similar as in
$\epem$ annihilation --- from hadronic jet production and event shapes.
Improved QCD predictions as well as new experimental studies 
provided new results from a combined study of world data on
structure functions, and from jet production at HERA.

\subsubsection{$\as$ from world data on non-singlet structure functions}

Physical processes in lepton-nucleon and
in hadron-hadron collisions depend on quark- and
gluon-densities in the nucleon.
Assuming factorisation between short-distance, hard scattering processes which
can be calculated using QCD perturbation theory, and low-energy or long-range
processes which are not accessible by perturbative methods,
such cross sections are 
parametrised by a set of structure functions $F_i$ ($i$= 1,2,3).

The energy dependence of structure functions is given by perturbative QCD.
A study \cite{bluemlein} of the available world data 
on deep inelastic lepton-proton and lepton-deuteron scattering provided
a determination of the valence quark parton densities and of
$\as$ in wide ranges of the Bjorken scaling variable $x$ and $Q^2$.
In the non-singlet case, where heavy flavour effects are negligibly small, 
the analysis is extended to 4-loop level, i.e. to QCD in
N3LO perturbative expansion.

The determination of $\as$ to this level results in
$$\amz = 0.1142 \pm 0.0023 \ , $$
where the total error includes a theoretical uncertainty of $\pm 0.0008$
which is taken from the difference between the N3LO and the NNLO result.
This value will be included in the determination of the world average
of $\amz$.

As it appears, fits of $\as$ in determinations of parton density
functions from purely deep inelastic scattering processes
result in somewhat smaller values than those which also include
hadron collider data, see e.g. \cite{mst} obtaining 
$\amz = 0.1171 \pm 0.0037$. 
They also are systematically smaller than the
world average value of $\amz$, see below.

\subsubsection{$\as$ from jet production in deep inelastic scattering processes}

Measruements of $\as$ from jet production in deep inelastic lepton-
nucleon scattering at the HERA collider have been and continue
to be an active field of research.
Inclusive as well as differential jet production rates were studied in
the energy range of $Q^2\sim 10$ up to 15000~GeV$^2$, based on  similar
jet definitions and algorithms as used in $\epem$ annihilation. 

In a recent summary and combination \cite{glasman} of precision 
measurements at HERA, values of $\as$ where determined from 
fits of NLO QCD predictions to data of inclusive
jet cross sections in neutral current deep inelastic scattering at 
high $Q^2$  \cite{zeus-jets,h1-jets}.
The overall combined result,
$$\amz = 0.1198 \pm 0.0032 \ ,$$
has a reduced theoretical uncertainty of $\pm 0.0026$ (added in 
quadrature to the experimental error of $\pm 0.0019$)
compared to previous combinations, due to carefully selected 
ranges of data in $Q^2$ ranges where theoretical uncertainties
are minimal \cite{glasman}.
This combined result will be included in the determination
of the world average of $\amz$.

\subsection{$\as$ from hadronic event shapes and jet production
in $\epem$ annihilation}

Observables parametrizing hadronic event shapes and jet production rates 
are the classical inputs for $\as$ studies in $\epem$ annihilation. 
The measurements summarised in previous reviews 
\cite{concise,biebel,kluth,sb-06}
were based on QCD predictions in NLO, which partly included summation
of next-to-leading logarithms (NLLA) to all orders.
As one of the most notable and long awaited theoretical improvements,
complete NNLO predictions became available recently
\cite{nnlo-shapes}, which are also matched with 
leading and next-to-leading logarithms resummed to all orders
\cite{nnlo-nlla} (NNLO + NLLA).

The advancement in theoretical descriptions was instantly used to 
determine $\as$ from data of previous $\epem$ annihilation experiments,
from the PETRA and the LEP colliders which operated from
1979 to 1986 and from 1989 to 2000, respectively.
The usage of data of past experiments demonstrates the need to
preserve data as well as reconstruction-, simulation- and analysis-software
for a time-span 
significantly exceeding the usual $\sim 5$-year period of post-data 
taking analysis.

A re-analysis \cite{dissertori} of the ALEPH data from LEP, in the
c.m. energy range from 90 to 206~GeV, based on six event shape and
jet production observables, results in
$$ \amz = 0.1224 \pm 0.0039 \ .$$
The total error contains an experimental uncertainty of
0.0013 and a theoretical uncertainty, mainly from hadronisation 
and from renormalisation scale dependences, of 0.0037.
This result is obtained using NNLO+NLLA QCD predictions; in
NNLO alone, the central value is slightly higher (0.1240) and the
total error is sligthly smaller (0.0032). 
NNLA terms, although 
they should provide a more complete perturbation series, 
tend to introduce somewhat larger scale uncertainties \cite{dissertori}.

Similar results are available from a re-analysis of data from
the JADE experiment at PETRA \cite{jade}, from six event shape and
jet observables at six c.m. energies in the c.m. energy range
from 14 to 46~GeV:
$$\amz = 0.1172 \pm 0.0051 \ .$$
The total error contains an experimental uncertainty of
0.0020 and a theoretical uncertainty of 0.0046.
Also this result is obtained using QCD predictions in NNLO+NLLA;
the value for NNLO alone is $\amz = 0.1212  \pm 0.0060$.

Both the NNLO+NLLA results from ALEPH and from JADE data are retained
for the determination of the new world average value of $\amz$ in this
review. 
Because they are based on data at different c.m. energy ranges and
from two independent experiments, they add valuable and independent
information not only on the world average, but also on the
experimental verification of the running of $\as$.
These two results of $\amz$ are included in figure~\ref{fig:asmz-09}; 
the respective values of $\as (Q)$, obtained at different
values of c.m. energies, are displayed in figure~\ref{fig:asq-09}.

Recently, the event shape observable thrust \cite{thrust} was also
studied from LEP data using methods of effective field theory 
\cite{fieldtheory}. 
Starting from a factorisation theorem in soft-collinear effective
theory, the thrust distribution is calculated including resummation
of the next-to-next-to-next-to-leading logarithms (N3LL).
The result of this analysis is $\amz = 0.1172 \pm 0.0021$, whereby the
error includes a theoretical uncertainty of $\pm 0.0017$.
Although this is formally one of the smallest errors quoted on
measurements of $\amz$, this result is not 
explicitly included in the world average
calculated below.
The reason for this decision is two-fold:
First, the LEP thrust data are already included in the 
re-analysis of ALEPH data described above.
Second, this analysis based on effective field theory, although being a
highly interesting, alternative approach to studies 
based on standard perturbation
theory, is not yet in a state of comparable reliability because it
is based on one event shape observable only, and therefore misses an 
important verification of potential systematic uncertainties.

\subsection{$\as$ from electroweak precision data}
The determination of $\as$ from totally inclusive observables,
like the hadronic width of the $\tau$ lepton discussed above, 
or the total hadronic decay width of the $Z^0$ boson, are
of utmost importance because they lack many sources of systematic uncertainties, experimental as well as theoretical, which differential
distributions like event shapes or jet rates suffer from.
In this sense, the ratio of
the hadronic to the leptonic partial decay width,
$R_Z = \Gamma (Z^0 \rightarrow hadrons) / \Gamma (Z^0 \rightarrow
 \epem )$, is a \oq gold plated" observable.

Since 1994, the QCD correction to $R_Z$ is known in NNLO QCD
\cite{rznnlo}. 
The measured value from LEP, $R_Z = 20.767 \pm 0.025$ \cite{lepew-0404},
results in $\amz = 0.1226 \pm 0.0038$, where the error is experimental.
An additional theoretical uncertainty was estimated \cite{concise} as
$^{+0.0043}_{-0.0005}$.

As already mentioned, the full N3LO prediction of $R_Z$, i.e. in $\oaaaa$
perturbative expansion, is now available \cite{n3lo}.
The negative $\oaaaa$ term results in an increase of $\amz$
by 0.0005, such that the actual result from the measured value
of $R_Z$ is: $\amz = 0.1231 \pm 0.0038 \ .$
Defining the remaining theoretical uncertainty as the difference
between the NNLO and the N3LO result, the theory error would not visibly
contribute any more, given the current size of the experimental error
on $\amz$ of $\pm 0.0038$.

A more precise value of $\amz$ can be obtained from general fits to all
existing electro-weak precision data, using data from the LEP and the SLC
$\epem$ colliders as well as measurements of the top-quark mass and
limits on the Higgs boson mass from Tevatron and LEP.
Such global fits result in values of $\amz$ with reduced 
experimental errors. These values, however,
were consistently smaller than (but still compatible with) 
the ones obtained from
$R_Z$ alone, see e.g. \cite{lepew-0404}.

A recent revision of the global fit to electroweak precision data \cite{hoecker}, 
based on a new generic fitting package $Gfitter$ \cite{gfitter},
on the up-to-date QCD corrections in N3LO,
on proper inclusion of the current limits from direct
Higgs-searches at LEP and at the Tevatron
and on other improved details, results in
$$\amz = 0.1193^{+0.0028}_{-0.0027}\pm 0.0005 \ ,$$
where the first error is experimental and the second is theoretical,
estimated by the difference of the results in NNLO and in N3LO QCD.

\renewcommand{\arraystretch}{1.2}
\begin{table*}[ht]
{
\caption{
Summary of recent measurements of $\amz$.
All eight measurements will be included in 
determining the world average value of $\amz$. 
The rightmost two columns give the exclusive mean value of
$\amz$ calculated {\em without} that particular measurement,
and the number of standard deviations between this
measurement and the respective exclusive mean,
treating errors as described in the text.
The inclusive average from {\em all} listed measurements
gives $\amz = 0.11842 \pm 0.00067$.
\label{tab:assel}}
\begin{center}
\begin{tabular}{|l|c|c|c||c|c|}
   \hline 
Process & Q [GeV] & $\as (Q)$ & $\amz$ & excl. mean $\amz$ & std. dev.\\
\hline 
$\tau$-decays        & 1.78     & $0.330\pm0.014$ & $0.1197 \pm 0.0016$              
 & $0.11818 \pm 0.00070$ & 0.9 \\
DIS [$F_2$] & 2 - 15 & -- & $0.1142 \pm 0.0023$              
 & $0.11876 \pm 0.00123$ & 1.7 \\
DIS [e-p $\rightarrow$ jets] & 6 - 100 & -- & $0.1198 \pm 0.0032$       
 & $0.11836 \pm 0.00069$ & 0.4 \\
${\rm Q\overline{Q}}$ states & 7.5 & $0.1923\pm0.0024$ & $0.1183 \pm 0.0008$  & $0.11862 \pm 0.00114$ & 0.2 \\
$\Upsilon$ decays & 9.46 & $0.184^{+0.015}_{-0.014}$ & $0.119^{+0.006}_{-0.005}$ &$ 0.11841 \pm 0.00070$ & 0.1 \\
$\epem$ [jets \& shps] & 14 - 44 & -- & $0.1172\pm0.0051$ 
  & $0.11844 \pm 0.00076$ & 0.2 \\
$\epem$ [ew] & 91.2 & $0.1193 \pm 0.0028$ & $0.1193 \pm 0.0028$ 
 & $0.11837 \pm 0.00076$ & 0.3 \\
$\epem$ [jets \& shps] & 91 - 208  & -- & $0.1224 \pm 0.0039$               
 & $0.11831 \pm 0.00091$ & 1.0 \\ \hline
\hline
\end{tabular}
\end{center}
}
\end{table*}

\section{The 2009 world average of $\amz$}

\begin{figure}[ht]
\resizebox{0.49\textwidth}{!}{%
  \includegraphics{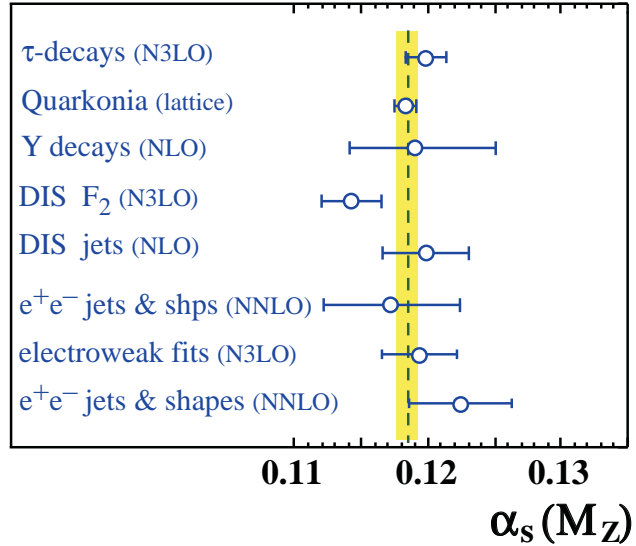}
}
\caption{
Summary of measurements of $\amz$.
The vertical line and shaded band 
mark the final world average value 
of $\amz = 0.1184 \pm 0.0007$ determined
from these measurements.
\label{fig:asmz-09}}
\end{figure}

The new results discussed in the previous section are summarised 
in table~1 and in figure~\ref{fig:asmz-09}.
Since all of them are based on improved theoretical predictions and
methods, and/or
on improved data quality and statistics, they supersede and replace their respective precursor results which were summarised in a previous
review \cite{sb-06}.
While those previous results continue to be valid measurements,
they are not discussed in this review again, and they will not be 
included in the determination of a combined
world average values of $\amz$, according to the following reasons:

\begin{enumerate}

\item
from each class of measurements, only the most advanced
and complete analyses shall be included in the new world average; 

\item
older measurements $not$ being complemented or superseded by the most
recent results listed above, as e.g. results from sum rules, 
from singlet structure functions of deep inelastic scattering,
and from jet production and b-quark production at hadron colliders,
are not
included because their relatively large overall uncertainties,
in general, will not give them a sizable weight but will complicate
the definition of the overall error of the combined value of $\amz$; 

\item
restricting the new world average to the  most recent and most
complete (i.e. precise) results allows to examine the consistency
between the newest and the previous generations of measurements
and reviews.
\end{enumerate}

\subsection{Numerical procedure}

The average
$\overline{x}$ of a set of $n$ different, uncorrelated measurements 
$x_i$ of a particular quantity $x$ 
with individual errors or uncertainties, $\Delta x_i$
is commonly defined using the method of $least$ $squares$
(see e.g. \cite{pdg}: 
For $x_i$ being independent
and statistically distributed measures with a common expectation
value $\overline{x}$ but with different variances $\Delta x_i$, the
weighted average is defined by
\begin{equation} \label{uncorr}
\overline{x} = \frac{\sum_{i=1}^{n} w_i x_i}{\sum_{i=1}^{n} w_i} \,
\end{equation}
and the variance $\Delta \overline{x}$ of $\overline{x}$ is minimised by
choosing
\begin{equation} \label{uncorrerr}
\Delta \overline{x}^2 = \frac{1}{\sum_{i=1}^{n} \frac{1}{\Delta
   x_i^2}}\ , \ \textrm{i.e.}\  
w_i = \frac{1}{\Delta x_i^2} \ .
\end{equation}
The quality of the average is defined by the $\chi^2$ variable,
\begin{equation} \label{uncorrchi}
\chi^2 = \sum_{i=1}^{n} \frac{(x_i - \overline{x})^2}{\Delta x_i^2}\ 
\end{equation}
which is, for uncorrelated data, expected to be equal to the number 
of degrees of freedom, $n_{df}$:
$$ \chi^2 = n_{df} = n - 1 \ .$$

The results summarised in table~1, however, 
are not independent of each other.
They are, in the most general sense, correlated to an unknown degree.
While the statistical errors of the data and the experimental systematic
uncertainties contained in the errors are 
independent and uncorrelated, the theoretical uncertainties are
very likely to be (partly) correlated between different results,
because they are all based on applying the same underlying theory,
i.e. QCD, and similar methods to obtain estimates of theoretical
uncertainties are being used. 

For some observables, like e.g. the hadronic widths of the
$Z^0$ boson and the $\tau$ lepton, the theoretical predictions and hence,
their uncertainties, are known to be correlated by almost 100\%.
For other cases, like the results based on lattice QCD and those
based on QCD perturbation theory, it can be assumed that their
theoretical uncertainties are not correlated at all.
In addition to the inherent lack of knowledge of theoretical correlations,
estimates of theoretical uncertainties, in general, are performed in
widely different ways, using different methods 
and different ranges of parameters. 

The presence of correlated errors, if using the 
equations given above, is usually signalled by 
$\chi^2 < n_{df}$.
Values of $\chi^2 > n_{df}$, in most practical cases, 
are a sign of possibly underestimated errors.
In this review, both these cases are pragmatically
handled in the following way:

In the presence of correlated errors, described
by a covariance matrix $C$, the optimal procedure to determine the
average $\overline{x}$ is to minimise the $\chi^2$ function
$$ \chi^2 = \sum_{i, j = 1}^{n} (x_i - \overline{x})(C^{-1})_{ij}
   (x_j - \overline{x})\ ,$$
which leads to 
$$\overline{x} = \left( \sum_{ij} (C^{-1})_{ij} x_j \right)
  \left( \sum_{ij} (C^{-1})_{ij} \right)^{-1} $$
and
$$\Delta \overline{x}^2 = \left( \sum_{ij} (C^{-1})_{ij} \right)^{-1} \ .$$

The choice of $C_{ii} = \Delta x_{i}^2$ and  
$C_{ij} = 0$ for $i \neq j$ retains the uncorrelated case
given above. 
In the presence of correlations, however, the resulting
$\chi^2$ will be less than $n_{df} = n - 1$.
In order to allow for an unknown $common$ degree of a correlation
$f$, the method proposed in \cite{schmelling} will be applied by
choosing $ C_{ij} = f \times \Delta x_i \times \Delta x_j$ and
adjusting $f$ such that $\chi^2 = n - 1$.

For cases where the uncorrelated error determimation results in
$\chi^2 > n_{df}$, and in the absence of knowledge which of
the errors $\Delta x_i$ are possibly underestimated,
all individual errors are scaled up by a common factor $g$ such that
the resulting value of $\chi^2 / n_{df}$, using the definition for
uncorrelated errors, will equal unity.

Note that both for values of $f > 0$ or $g > 1$, 
$\Delta \overline{x}$ increases,
compared to the uncorrelated ($f = 0$ and $g = 1$) case.

\subsection{Determination of the world average}

The eight different determinations of $\amz$ summarised and discussed
in the previous section are listed in table~1 and
are graphically displayed in figure~\ref{fig:asmz-09}.
Applying equations \ref{uncorr}, \ref{uncorrerr} and \ref{uncorrchi} to
this set of measurements, assuming that the errors are not
correlated, results in an average value of
$\amz = 0.11842 \pm 0.00063$ with $\chi^2 / n_{df} = 5.4 / 7$.

The fact that $\chi^2 < n_{df}$ signals a possible correlation
between all or subsets of the eight input results.
Assuming an overall correletion factor $f$ and
demanding that $\chi^2 = n_{df} = 7$
requires $f = 0.23$, inflating the overall error from 0.00063 to
0.00089.

In fact, there are two pairs of results which are known to be largely
correlated:
\begin{itemize}
\item 
the two results from $\epem$ event shapes based on the data from JADE 
and from ALEPH use the same theoretical
predictions and similar hadronisation models to correct these predictions
for the transitions of quarks and gluons to hadrons.
While the experimental errors are uncorrelated, the theoretical 
uncertainties may be assumed to be correlated to 100\%.
The latter accounts for about 2/3 to 3/4 of the total errors.
An appropriate choice of correlation factor between the two may then
be $f = 0.67$.
\item
the QCD predictions for the hadronic widths of the $\tau$-lepton and the
$Z^0$ boson are essentially identical, so the respective results on
$\as$ are correlated, too.
The values and total errors of $\amz$ from $\tau$ decays must therefore be
correlated to a large extend, too.
In this case, however, the error of one measurement is almost entirely 
determined by the experimental error ($Z^0$-decays), while the other,
from $\tau$-decays, is mostly theoretical. 
A suitable choice of the correlation factor between both these results may 
thus be $f = 0.5$.
\end{itemize}

Inserting these two pairs of correlations into the error matrix $C$,
the $\chi^2 / n_{df}$ of the averaging procedure results in  
6.8/7, and the overall error on the (unchanged) central value of
$\amz$ changes from 0.00063 to 0.00067.
Therefore the new world average value of $\amz$ is defined to be
$$\fbox{$\amz = 0.1184 \pm 0.0007$.}$$

For seven out of the eight measurements of $\amz$, the average value
of 0.1184 is within one standard deviation of their assigned
errors. 
One of the measurements, from structure functions \cite{bluemlein}, 
deviates from
the mean value by more than one standard deviation, 
see figure~\ref{fig:asmz-09}.

The mean value of $\amz$ is potentially dominated by the
$\as$ result with the smallest overall assigned uncertainty, which is
the one based on lattice QCD \cite{davies08}.
In order to verify this degree of dominance on the average result
and its error, and to test the compatibility of each of the
measurements with the others,
exclusive averages, leaving out one of the
8 measurements at a time, are calculated.
These are presented in the $5^{th}$ column of table~1,
together with the corresponding number of
standard deviations
\footnote{
The number of standard deviations is defined as the square-root of the
value of $\chi^2$.}
between the
exclusive mean and the respective single measurement.

As can be seen, the values of exclusive means vary only
between a minimum of 0.11818 and a maximum 0.11876.
Note that in the case of these exclusive means and according to
the "rules" of calculating their overall errors, in four out of the
eight cases small error scaling factors of $g = 1.06 ... 1.08$ had
to be applied, while in the other cases, overall correlation
factors of about 0.1, and in one case of 0.7, had to be applied to
assure $\chi^2 / n_{df} = 1$.
Most notably, the average value $\amz$ changes to
$\amz = 0.1186 \pm 0.0011$ when omitting the result from lattice QCD.

\section{Summary and Discussion}

In this review, new results and measurements of $\as$ are summarised, and
the world average value of $\amz$, as previously given in 
\cite{concise,sb-06,pdg}, is updated. 
Based on eight recent measurements, which partly use new and 
improved N3LO, NNLO and lattice QCD predictions, the new average
value is 
$$\amz = 0.1184 \pm 0.0007\ , $$
which corresponds to
$$\lamsb^{(5)} = \left( 213 \pm 9\ \right) \textrm{MeV} \ .$$
This result is consistent with the one obtained in the
previuos review three years ago \cite{sb-06}, 
which was $\amz = 0.1189 \pm 0.0010$.
The previous and the actual world average have been obtained from 
a non-overlapping set of single results; their agreement therefore 
demonstrates a large degree of compatibility between the old and the
new, largely improved set of measurements.

The individual mesurements, as listed in table~1 and displayed in
figure~\ref{fig:asmz-09},
show a very satisfactory agreement with each other and with the
overall average:
only one out of eight measurements exceeds a deviation from the average by 
more than one standard deviation, and the largest deviation between any two
out of the eight results, namely the ones from $\tau$ decays and from
structure functions, amounts to 2 standard deviations
\footnote{assuming their assigned total errors to be fully uncorrelated.}.

There remains, however, an apparent and long-standing systematic 
difference:
results from structure functions prefer smaller values of $\amz$ than
most of the others, i.e. those from $\epem$ annihilations, from 
$\tau$ decays, but also those from jet production in
deep inelastic scattering.
This issue apparently remains to be true, 
although almost all of the new results are
based on significantly improved QCD predictions, up to N3LO for structure functions, $\tau$ and $Z^0$ hadronic widths, and NNLO for $\epem$
event shapes.

The reliability of \oq measurements" of $\as$ based on \oq experiments"
on the lattice have gradually improved over the years, too.
Including vaccum polarisation of three light quark flavours
and extended means to understand and correct for finite lattice spacing
and volume effects, the overall error of these results 
significally decreased over time, while the value of $\amz$ 
gradually approached the world average.
Lattice results today quote the smallest overall error on $\amz$;
it is, however, ensuring to see and note that the world average
$without$ lattice results is only marginally different, while the
small size of the total uncertainty on the world average is,
naturally, largely influenced by the lattice result.

\begin{figure}[ht]
\resizebox{0.49\textwidth}{!}{%
\includegraphics{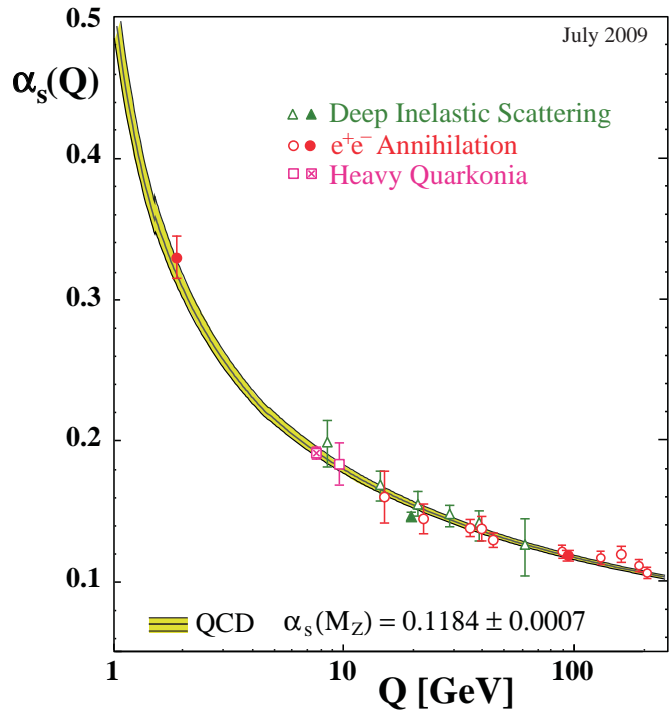}
}
\caption{
Summary of measurements of $\as$ as a function of the
respective energy scale $Q$.
The curves are QCD predictions for the combined world
average value of $\amz$, in 4-loop approximation and using 3-loop
threshold matching at the heavy quark pole masses
$M_c = 1.5$~GeV and $M_b = 4.7$~GeV.
Full symbols are results based on N3LO QCD,
open circles are based on NNLO, open triangles and squares on
NLO QCD. 
The cross-filled square is based on lattice QCD.
The filled triangle at $Q = 20$~GeV (from DIS structure functions)
is calculated 
from the original result which includes
data in the energy range from $Q =$2 to 170 GeV.
\label{fig:asq-09}}
\end{figure}

In order to demonstrate the agreement of measurements with the
specific energy dependence of $\as$ predicted by QCD,
in figure~\ref{fig:asq-09}  the recent measurements of $\as$ 
are shown as
a function of the energy scale $Q$.
For those results which are based on several $\as$ determinations
at different values of energy scales $Q$, the individual
values of $\as(Q)$ are displayed.
For the value from structure functions
such a breakup is not possible; instead, the corresponding result
derived for a 
typical energy scale of $Q = 20$~GeV is displayed.

\begin{figure}[ht]
\resizebox{0.49\textwidth}{!}{%
  \includegraphics{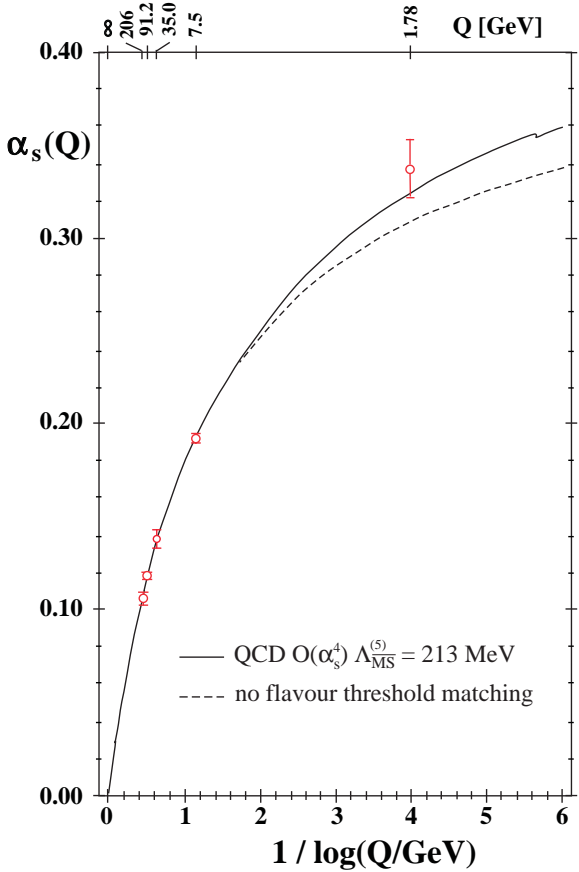}
}
\caption{
Selected measurements of $\as$,
as a function of the inverse logarithm of
the energy scale $Q$, in order to demonstrate
concordance with Asymptotic Freedom.
The full line is the QCD prediction in 4-loop approximation
with 3-loop threshold matching at the heavy quark pole masses.
The dashed line indicates extrapolation of the 5-flavour
prediction without threshold matching.
\label{fig:as09-logq}}
\end{figure}

The measurements significantly prove the validity of the
concept of asymptotic freedom; they are in perfect agreement
with the QCD prediction of the running coupling.
This is further corroborated by figure~\ref{fig:as09-logq},
where a selected sample of the measurements is plotted, now
as a function of $1 / \log{Q}$, in order to demonstrate the
data reproducing the specific logarithmic shape of the running
as predicted by QCD, signalling that indeed
$\as (Q) \rightarrow 0$ for $Q \rightarrow \infty$.

What are the future prospects of measurements of $\as$?
With the given degree of data and theory precision, further
improvements will be difficult and may take quite some time.
Experimentally,
a linear $\epem$ collider, especially if run in the \oq Giga-Z" mode,
has the potential to decrease the dominating experimental error of $\amz$ 
from the measurement of $R_Z$, down and below its theoretical
uncertainty which currently is assumed to be, in N3LO, $\pm 0.0005$.

While it is unlikely that QCD perturbation theory will
improve to yet one order higher than the existing N3LO or NNLO
predictions, improvements are likely, and actually are very neccessary, 
for QCD predictions of jet production in deep inelastic scattering
and in hadron collisions, where calculations currently are limited to
NLO.
The precision of QCD tests, but also the sensitivity for observing new physics signals at the LHC, will largely depend on a further advancement of
QCD predictions for hadron collisions.

Future improvement of theoretical predictions and models require
the conservation of data and of reconstruction and simulation code
of current and past experiments;
especially in the case of deep inelastic scattering data,
re-application of improved predictions and models carry a large
potential for future advancements in this field.

Last but not least,
further developments of non- perturbative methods are mandatory 
to bridge the gap
between quarks and gluons and their final states, hadrons.  
They may in fact shed more light into the systematic
differences between some classes of measurements as discussed above.


\begin{thebibliography}{19}
%
\bibitem{qcd}
H. Fritzsch, M. Gell-Mann and H. Leutwyler
Phys. Lett. \textbf{B47}, (1973) 365; \\
D.J. Gross, F. Wilczek, Phys. Rev. Lett. \textbf{30} (1973) 1343; 
Phys. Rev. \textbf{D8} (1973) 3633;\\
H.D. Politzer, Phys. Rev. Lett. \textbf{30} (1973) 1346.
%
\bibitem{ellis-book}
R.K. Ellis, W.J. Stirling and B.R. Webber, {\it QCD and Collider Physics},
Cambridge University Press, 1996.
%
\bibitem{collins-book}
J.C. Collins, {\it Renormalization}, Cambridge University Press, 1984.
%
\bibitem{yndurain-book}
F.J. Yndurain, {\it The Theory of Quark and Gluon Interactions}, 
Springer-Verlag, 1999.
%
\bibitem{dissertori-book}
G. Dissertori, I.G. Knowles, M. Schmelling,
\textit{High energy experiments and theory}, 
Oxford, UK: Clarendon (2003) 538 p, (International series of monographs on physics. 115).
%
\bibitem{pdg}
PDG, C. Amsler et al., Phys. Lett. \textbf{B667} (2008).
%
\bibitem{concise}
S. Bethke, J. Phys. \textbf{G26}, (2000) R27;
hep-ex/0004021.
%
\bibitem{beta4loop}
T. van Ritbergen, J.A.M. Vermaseren, S.A. Larin, Phys. Lett. 
\textbf{B400} (1997) 379. \\
M. Czakon, Nucl.Phys. \textbf{B710} (2005) 485; 
hep-ph/0411261.
%
\bibitem{alphas-4loop}
K.G. Chetyrkin, B.A. Kniehl and M. Steinhauser, Phys. Rev. Lett. 
\textbf{79} (1997)
2184.
%
\bibitem{bernreuther}
W. Bernreuther and W. Wetzel, Nucl. Phys. \textbf{B197} (1982) 128.
%
\bibitem{larin}
S.A. Larin, T. van Ritbergen and J.A.M. Vermaseren, Nucl. Phys. \textbf{B438} (1995) 278.
%
\bibitem{nnlo-shapes}
A. Gehrmann-de Ridder et al., JHEP 0712 (2007) 094; 
arXiv:0711.4711 [hep-ph]. \\
S. Weinzierl, Phys. Rev. Lett. \textbf{101} (2008) 162001;
arXiv:0807.3241 [hep-ph].
%
\bibitem{n3lo}
P.A. Baikov, K.G. Chetyrkin, J.H. K\"uhn, Phys. Rev. Lett.
\textbf{101} (2008) 012002, arXiv:0801.1821 [hep-ph].
%
\bibitem{resummation}
S. Catani, L. Trentadue, G. Turnock, B.R. Webber, Nucl. Phys. 
\textbf{B407} (1993) 3.
%
\bibitem{stevenson}
P.M. Stevenson, Phys. Rev. \textbf{D23} (1981) 2916.
%
\bibitem{grunberg}
G. Grunberg, Phys. Rev. \textbf{D29} (1984) 2315.
%
\bibitem{blm}
S.J. Brodsky, G.P. Lepage and P.B. Mackenzie, Phys. Rev. \textbf{D28}
(1983) 228.
%
\bibitem{siggiscale}
S. Bethke, Z. Phys \textbf{C43} (1989) 331.
%
\bibitem{msbar}
W.A. Bardeen et al., Phys. Rev. \textbf{D18} (1978) 3998.
%
\bibitem{string}
T. Sjostrand, Comput. Phys. Commun. \textbf{27} (1982) 243.
%
\bibitem{pythia}
T. Sjostrand, S. Mrenna and P Skands, 
Comput.Phys.Commun. \textbf{178} (2008) 852; 
arXiv:0710.3820 [hep-ph].
%
\bibitem{herwig}
G. Marchesini and B.R. Webber, Nucl.Phys. \textbf{B238} (1984) 1; \\
G. Corcella et al., JHEP \textbf{0101} (2001) 010,  
hep-ph/0011363.
%
\bibitem{powcor}
Yu.L. Dokshitzer, B.R. Webber, Phys. Lett. \textbf{B352} (1995) 451; \\
Yu.L. Dokshitzer, G. Marchesini, B.R. Webber, Nucl.Phys. \textbf{B469} (1996) 93; \\
Yu.L. Dokshitzer, B.R. Webber, Phys. Lett. \textbf{B404} (1997) 321; \\
S. Catani, B.R. Webber, Phys. Lett. \textbf{B427}(1998) 377; \\
Yu.L. Dokshitzer, A. Lucenti, G. Marchesini, G.P. Salam,
Nucl.Phys. \textbf{B511} (1998) 296; JHEP \textbf{05} (1998) 003.
%
\bibitem{weisz}
P. Weisz, Nucl.Phys. B (Proc. Suppl.) \textbf{47} (1996) 71; hep-lat/9511017.
%
\bibitem{fodor}
S. Durr et al., Science \textbf{322} (2008) 1224; 
arxiv:0906.3599 [hep-lat].
%
\bibitem{davies08}
C.T.H. Davies et al., HPQCD Collab., Phys.Rev. \textbf{D78} (2008) 114507; 
arXiv:0807.1687 [hep-lat].
%
\bibitem{altarelli}
G. Altarelli, Ann.Rev.Nucl.Part.Sci. \textbf{39} (1989) 357.
%
%
\bibitem{sb-06}
S. Bethke, Prog.Part.Nucl.Phys. \textbf{58} (2007) 351; 
hep-ex/0606035.
%
%
%
\bibitem{braaten}
E. Braaten, S. Narison and A. Pich, Nucl. Phys.  \textbf{B373} (1992) 581.
%
\bibitem{ew1}
W. Marciano and A. Sirlin, Phys. Rev. Lett. \textbf{61} (1988) 1815.
%
\bibitem{ew2}
E. Braaten and C.S. Li, Phys. Rev. \textbf{D42} (1990) 3888.
%
\bibitem{ope}
M.A. Shifman, L.A. Vainshtein, V.I. Zakharov, Nucl. Phys. 
\textbf{B147} (1979) 385.
%
\bibitem{davier06}
M. Davier, A. H\"ocker, Z. Zhang, Rev.Mod.Phys. \textbf{78} (2006) 1043; 
hep-ph/0507078.
%
\bibitem{beneke}
M. Beneke and M. Jamin, Martin Beneke, JHEP \textbf{0809} (2008) 044; 
arXiv:0806.3156 [hep-ph].
%
\bibitem{davier08}
M. Davier et al., Eur.Phys.J. \textbf{C56} (2008) 305; 
arXiv:0803.0979 [hep-ph].
%
\bibitem{maltman}
K. Maltman, T. Yavin,  Phys.Rev. \textbf{D78} (2008) 094020; 
arXiv:0807.0650 [hep-ph].
%
\bibitem{menke}
S. Menke, arXiv:0904.1796 [hep-ph].
%
\bibitem{narison}
S. Narison, Phys.Lett. \textbf{B673} (2009) 30; 
arXiv:0901.3823 [hep-ph]
%
\bibitem{lep-tau}
K. Ackerstaff et al., OPAL Collab., Eur. Phys. J. \textbf{C7} 
(1999) 571; hep-ex/9808019. \\
S. Schael et al., ALEPH Collab., Phys.Rept. \textbf{421} (2005) 191;
arXiv:hep-ex/0506072v1. 
%
\bibitem{babar}
B. Aubert et al., BABAR Collab., Phys.Rev.Lett.\textbf{100} (2008);arXiv:0707.2981 [hep-ex].
%
%
%
\bibitem{hq-review}
N. Brambilla et al., CERN-2005-005, hep-ph/0412158.
%
\bibitem{brambilla}
N. Brambilla et al., Phys. Rev \textbf{D75} (2007) 074014; hep-ph/0702079.
%
\bibitem{cleo-U}
D, Besson et al., CLEO Collab., Phys. Rev. \textbf{74} (2006) 012003;
hep-ex/0512061.
%
%
%
%
\bibitem{yndurain-79}
A. Gonzales-Arroyo, C. Lopez and F.J. Yndurain, Nucl. Phys. 
\textbf{B153} (1979) 161.
%
\bibitem{bluemlein}
J. Bl\"umlein, H. B\"ottcher and A. Guffanti, Nucl. Phys. B \textrm{774}
(2007) 182; hep-ph/0607200.
%
\bibitem{mst}
A.D. Martin et al., Phys. Lett. \textbf{B652} (2007) 292; 
arXiv:0706.0459 [hep-ph],
arXiv:0905.3531 [hep-ph].
%
\bibitem{glasman}
C. Glasman, J.Phys.Conf.Ser. \textbf{110} (2008) 022013, 
arXiv:0709.4426.
%
\bibitem{zeus-jets}
S. Chekanov et al., ZEUS Collab., Phys.Lett. \textbf{B649} (2007) 12; 
hep-ex/0701039.
\bibitem{h1-jets}
A. Aktas et al., H1 Collab., Phys.Lett. \textbf{B653} (2007) 134; 
arXiv:0706.3722 [hep-ex]. \\
F.D. Aaron et al., H1 Collab., arXiv:0904.3870 [hep-ex].
%
%
%
\bibitem{biebel}
O. Biebel, Phys. Rept. \textbf{340} (2001) 165.
%
\bibitem{kluth}
S. Kluth, Rept. Prog. Phys. \textbf{69} (2006) 1771;
hep-ex/0603011.
%
\bibitem{nnlo-nlla}
T. Gehrmann, G. Luisoni, H. Stenzel, Phys.Lett. \textbf{B664} (2008) 
265; arXiv:0803.0695. 
%
\bibitem{dissertori}
G. Dissertori et al., JHEP 0802 (2008) 040; 
arXiv:0712.0327 [hep-ph]. \\
G. Dissertori et al., arXiv:0906.3436 [hep-ph].
%
\bibitem{jade}
S. Bethke et al., arXiv:0810.1389 [hep-ex].
%
\bibitem{thrust} 
S. Brandt et al., Phys. Lett. \textbf{12} (1964) 57; \\
E. Farhi, Phys. Rev. Lett. \textbf{39}  (1977) 1587.
%
\bibitem{fieldtheory}
T. Becher and M.D. Scheartz, JHEP 0807 (2008) 034; 
arXiv:0803.0342 [hep-ph].
%
%
%
\bibitem{rznnlo}
S.A.Larin, T. van Ritbergen, J.A.M. Vermaseren, Phys. Lett. B320 (1994) 159; \\
K.G. Chetyrkin, O.V. Tarasov, Phys. Lett. B327 (1994) 114.
%
\bibitem{lepew-0404}
The LEP Collaborations ALEPH, DELPHI, L3 and OPAL; hep-ex/0509008.
%
\bibitem{hoecker}
H. Fl\"acher et al., Eur.Phys.J. \textbf{C60} (2009) 543; 
arXiv:0811.0009 [hep-ph]
%
\bibitem{gfitter}
http://cern.ch/Gfitter.
%
\bibitem{schmelling}
M. Schmelling, Phys. Scripta \textbf{51} (1995) 676. 
%


\end{thebibliography}
\end{document}